# Dust Formation in Astrophysical Environments: The Importance of Kinetics

A.G.G.M. Tielens [1,2]*

[1]Leiden Observatory, Leiden University, Leiden, Netherlands, [2]Astronomy Department, University of Maryland, College Park, MD, United States

Astronomical observations and analysis of stardust isolated from meteorites have revealed a highly diverse interstellar and circumstellar grain inventory, including a wide range of amorphous materials and crystalline compounds (silicates and carbon). This diversity reflects the wide range of stellar sources injecting solids into the interstellar medium each with its own physical characteristics such as density, temperature and elemental composition and highlights the importance of kinetics rather than thermodynamics in the formation of these compounds. Based upon the extensive literature on soot formation in terrestrial settings, detailed kinetic pathways have been identified for the formation of carbon dust in C-rich stellar ejecta. These have been incorporated in astronomical models for these environments. In recent years, the chemical routes in the nucleation of oxides and silicates have been the focus of much astronomical research. These aspects of stardust formation will be reviewed and "lessons" for dust formation in planetary atmospheres will be drawn with the emphasis on the influence of kinetics on the characteristics and structure of dust in these environments.

## 1 INTRODUCTION

Interstellar dust is an important component of the interstellar medium of galaxies. Dust regulates the radiative transfer of photons with energies less than the ionization energy of hydrogen (13.6 eV) and hence dominates the spectral energy distribution of galaxies as well as the heating of the gas in diffuse clouds and photodissociation regions. Dust will shield molecules against the destructive effects of penetrating UV photons and provides surfaces for active (catalytic) chemistry. Hence, dust plays an important role in setting the organic inventory of molecular clouds and regions of star and planet formation. Finally, dust locks up much of the heavy elements in solid form. These grains provide the building blocks from which pebbles, cometesimals, and planetesimals are formed in planet-forming disks around young stars. Clearly, the composition, origin and evolution of interstellar dust is a key question within astrophysics and textbooks on the interstellar medium generally devote several chapters to this topic (Tielens, 2005; Draine, 2011).

Direct information on the composition of interstellar dust is mainly derived from infrared spectroscopy. Broad 9.7 and 18 $\mu$m bands, dominating infrared extinction, are due to the stretching and bending vibrations of amorphous silicates. In many circumstellar environments (ejecta from asymptotic giant branch (AGB) stars and red supergiants, Herbig AeBe and T Tauri stars (Ultra) Luminous Infrared galaxies ((U)LIRGS)), the infrared signature of crystalline silicates and oxide grains are also present. Infrared absorption bands due to aliphatic (eg., $CH_2$ and $CH_3$ groups in $sp^3$

**TABLE 1 |** An inventory of circumstellar dust.

| Material | AGB | Post-AGB | PN | Nova | T tauri | Herbig AeBe | RSG | Wolf rayet | LBV | SN type II | Massive YSO |
|---|---|---|---|---|---|---|---|---|---|---|---|
| amorphous silicates | 1, 2 | 1 | 1 | 1 | 1 | 1 | 1 | | 1 | 1 | 1 |
| crystalline forsterite | 1, 2 | 1 | 1 | | 1 | 1 | 1 | | 1 | | |
| crystalline enstatite | 1, 2 | 1 | 1 | | 1 | 1 | 1 | | 1 | | |
| aluminum oxide | 1 (?), 2 | | | | | | | | | 2 | |
| Spinel | 1 (?), 2 | | | | | | | | | 2 | |
| TiO2 | 2 | | | | | | | | | | |
| Hibonite | 2 | | | | | | | | | | |
| MgO | 1 | | | | | | | | | | |
| Fe | 1 (?) | | | | | | | | | 1 | |
| PAHs | 1, 2 | 1 | 1 | 1 | | 1 | 1 | (1) | 1 | | |
| Amorphous carbon | 1 | 1 | 1 | 1 | | | | 1 | | | |
| Graphite | 2 | | | 2 | | | | | | 2 | |
| Diamond | | 1 | | | 1 | | | | | 2 | |
| SiC | 1, 2 | | 1 | 2 | | | | | | 2 | |
| other carbides | 2 | 1 (?) | | | | | | | | 2 | |
| Si3N4 | | | | | | | | | | 2 | |
| MgS | 1 | 1 | 1 | | | | | | 1 | | |
| carbonate | | 1 (?) | | 1 (?) | 1 (?) | | | | | | |
| Ice | 1 | 1 | 1 | | 1 | 1 | 1 | | | | 1 |

*Legend: 1: Astronomical data 2: Meteoritic data.*

*AGB: Low mass (< 8 $M_\odot$) stars on the Asymptotic Giant Branch. Post-AGB object: Low mass (< 8 $M_\odot$) stars in transition from the AGB phase to the Planetary nebula phase. PN: The white dwarf remaining after the phase of prodigious mass loss on the AGB ionizes the AGB-ejecta. The resulting glowing nebula is called a planetary nebula. Nova: The cataclysmic nuclear explosion caused by the accretion of hydrogen onto the surface of a white dwarf star can lead to ejection of material. T Tauri star: A low mass (~ 1 $M_\odot$ protostar). Herbig AeBe star: Intermediate-mass (1.5 < M < 10 $M_\odot$) pre main-sequence stars with spectral types A or B first recognized by Herbig. RSG (Red supergiant): Late and cool (T ~ 3000 K) stage in the evolution of massive stars (M > 8 $M_\odot$). Wolf Rayet star: Hot stars characterized by massive stellar winds. Some of these objects have C-rich composition and carbon dust condenses out in their ejecta. LBV (Luminous blue variable): The most massive, brightest and bluest stars are variable and may experience periods of eruptive mass loss (eg., η Car). SN type II: The explosion of a massive (M > 8 $M_\odot$) star at the end of its lifetime. Massive YSO: luminous and massive protostar characterized by vast amounts of cold dust and gas.*

bonded carbon) CH stretching and bending modes at 3.4 and 6.8 and 7.2 $\mu m$ attest to the presence of Hydrogenated Amorphous Carbon (HAC) dust. Aromatic compounds betray their presence through the Aromatic Infrared Bands at 3.3, 6.2, 7.7, 8.6, 11.2, and 12.7 $\mu m$ as well as through the strong 2,175 Å bump in the interstellar extinction curve. Finally, stardust isolated from meteorites provide a unique view of the composition of interstellar dust and, through their isotopic signature(s) a direct link to their stellar formation sites. **Table 1** summarizes the various compound identified. Each of the entries in this table is a story in itself, but space does not allow a detailed discussion. Instead the interested reader is referred to various reviews (Tielens et al., 1997; Zinner, 2003; Nittler & Ciesla, 2016), which summarize these discussions and provide references to the original literature. Astronomical identifications which are particularly ambiguous are labelled with (?) in this table.

Much of this dust is formed in the ejecta of stars, either in stellar winds associated with Asymptotic Giants Branch stars, red supergiants, carbon-rich Wolf-Rayet stars and Luminous Blue Variables or in explosive events such as supernovae and novae. As the gas cools down, chemistry will lead to the formation of molecules and small clusters that can serve as condensation nuclei for grain growth. There is a long history of studies on dust formation in stellar ejecta. While originally, these studies were built on classical nucleation theory and thermodynamics, in recent years, the field has evolved to a chemical approach of dust formation. With the advent of density functional theory in quantum chemistry, it has become feasible to calculate the energetics of small clusters and investigate the chemical routes involved in their growth. The first steps in this were set in the late-1980ies early-1990ies for carbon dust formation in AGB star ejecta, based upon the vast chemical literature on soot formation. In recent years, studies of oxide formation have also gone down this route. These developments can provide a useful framework for studies on dust formation in exoplanet atmospheres. Here, we will review the status of studies of dust formation in stellar ejecta as a guide for the latter.

# 2 THERMODYNAMICS VERSUS KINETICS

Thermodynamic studies of dust formation have a long history in astrophysics (Grossman and Larimer, 1974; Salpeter, 1977) tied to e.g., the mineral compounds identified in meteorites. Many of these compounds are high temperature condensates and their occurance is well described by thermodynamic calculations of molecular equilibria. During the earliest phases of accretion, the inner Solar nebula was very hot, and even the most resilient compounds will sublimate close to the star. As this material subsequently cools down, compounds will start to condense out sequentially. As densities are also very high, it is reasonable to describe the chemistry using chemical thermodynamics. The results of such calculations are generally summarized using the condensation temperature—the temperature at which 50% of an element has condensed out—at a given pressure and the detailed results of one such condensation calculation are shown in will depend on the (partial) pressures of the elements involved. These react on with the gases to form spinel ($MgAl_2O_4$), melilite ($Ca(AlMg)$ $(SiAl)_2O_7$) and diopside ($CaMg(SiO_3)_2$) at somewhat lower temperatures (1,450–1650 K). Diopside is converted into Ca-Al-Na silicates such as plagioclases (e.g., anorthite, $CaAl_2Si_2O_8$) and feldspar (e.g., albite $NaAlSi_3O_8$) around 1,000–1200 K. Only a minor fraction of the condensibles are involved in these steps. Major dust formation—around 1,400–1500 K—involves condensation of Fe, Mg and Si. Depending on pressure, this may start with metallic iron condensation followed by silicates or the other way around. In thermodynamic equilibrium, magnesium silicate condensation, starting with olivine (forsterite, $Mg_2SiO_4$), is the first major silicate condensation branch. Upon further cooling, forsterite will react on with SiO in the gas phase around 1300 K to form orthopyroxene (enstatite, $MgSiO_3$). The Mg-rich end members of olivine and pyroxene are converted into mixed-Mg-Fe compounds around 700 K. Upon further cooling, these models predict that metallic iron is converted into sulfides (∼ 700 K) and oxides, while the silicates react on with water to form phyllosilicates.

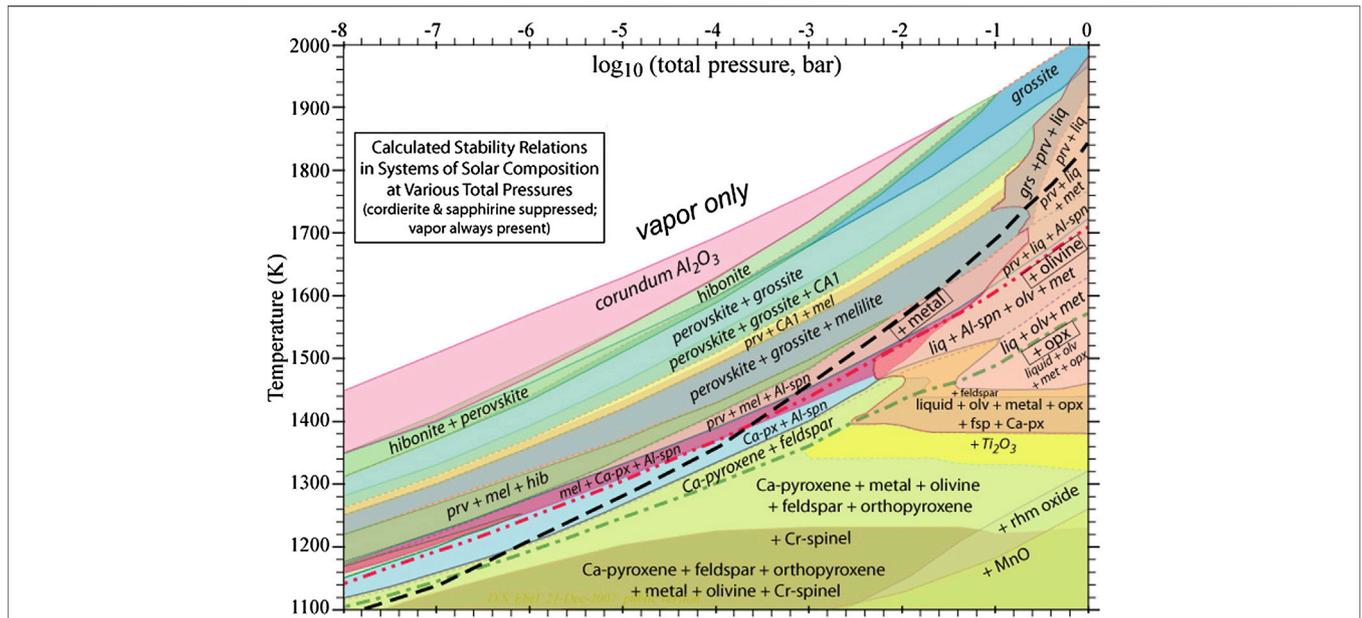

**FIGURE 1 |** Equilibrium stability regime of compounds in a nebula of solar elemental composition calculated using chemical thermodynamics. As the (inner) solar nebula cooled down from high temperatures, different compounds will condense out sequentially. The major condensates of metal, olivine, and orthopyroxene are indicated by dashed, dashed-dot-dot and dashed-dot lines, respectively. Figure taken from Ebel (2006).

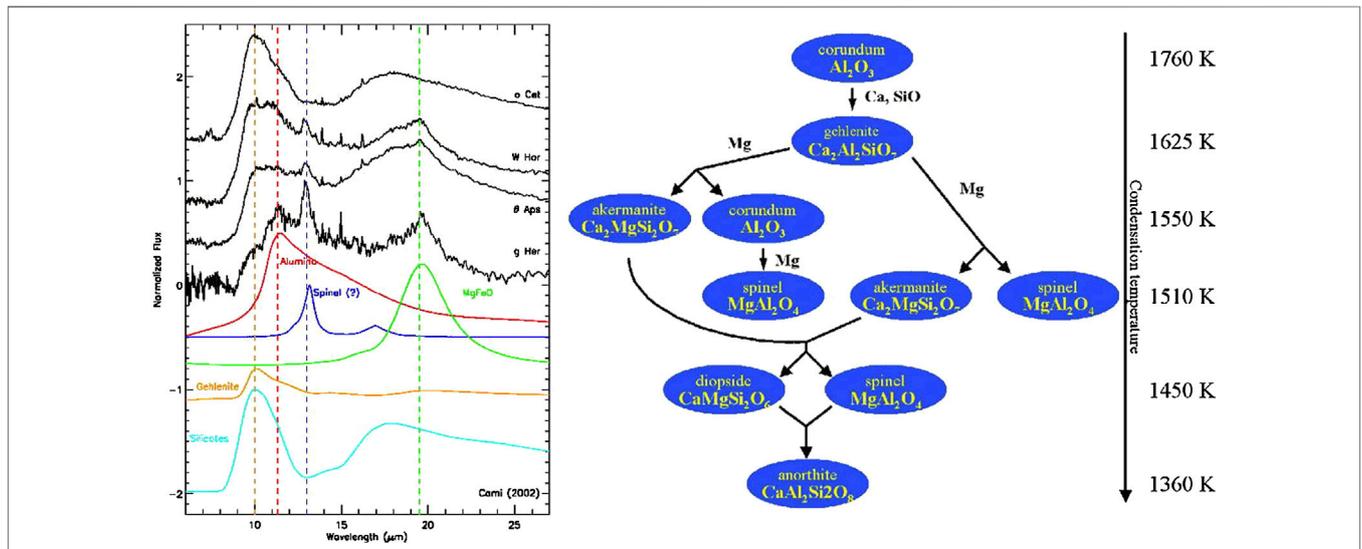

**FIGURE 2 |** Left: A selection of ISO SWS spectra of AGB stars illustrating the rich spectral diversity of the circumstellar dust compounds (Cami 2001). As the bottom traces of laboratory results demonstrate, these spectra show evidence for Aluminum oxides, spinel magnesium oxides, and amorphous (magnesium) silicates. Right: The condensation sequence of oxides and calcium-aluminum-silicates. The bulk of the dust is thought to condense out as magnesium silicates at 1250 K on top of these high temperature compounds. The objects in the left panel are thought to represent different stages in this dust condensation process "preserved" by freeze out. See text for more detail.

**Figure 1**. Condensation will start with the rare-earth elements such as Os, Zr, and Re and is then followed by oxide condensation. A simplified sequence for the oxides is shown in **Figure 2** and starts with aluminum and titanium oxides such as corundum ($Al_2O_3$) and perovskite ($CaTiO_3$) at about 1700 K. The exact temperature

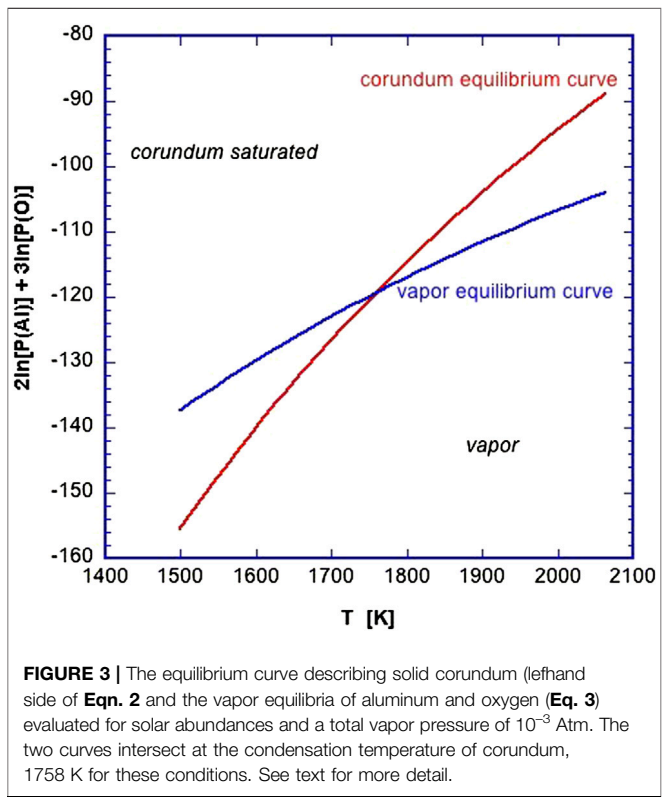

**FIGURE 3** | The equilibrium curve describing solid corundum (lefhand side of **Eqn. 2** and the vapor equilibria of aluminum and oxygen (**Eq. 3**) evaluated for solar abundances and a total vapor pressure of $10^{-3}$ Atm. The two curves intersect at the condensation temperature of corundum, 1758 K for these conditions. See text for more detail.

As an example, consider the condensation of aluminum oxide in thermodynamic equilibrium. The equilibrium between gaseous aluminum and oxygen and solid corundum can be written as,

$$2Al + 3O \leftrightarrow Al_2O_3, \tag{1}$$

This leads to,

$$-\ln\left[P_{Al}^2 P_O^3\right] = \ln\left[K_{eq}(Al_2O_3)\right] \tag{2}$$

Where the $P$ are the partial pressures of the compounds and $K$ is the equilibrium constant. Atomic Al is the dominant aluminum bearing species but oxygen is mainly locked up in $H_2O$. We can link the atomic O partial pressure to that of $H_2O$ and $H_2$ through the equilkibrium constants of $H_2$ and $H_2O$; viz.,

$$P_O = \frac{P_{H_2O}}{P_{H_2}} \frac{K_{eq}(H_2)}{K_{eq}(H_2O)}. \tag{3}$$

This can be solved graphically (**Figure 3**) where Gibbs free energies have been taken from compilations. The corundum condensation temperature is given by the intersection of the two curves. Above this temperature, aluminum will be in the gas phase and described by the (blue) vapor curve. Below that temperature, the (red) solid curve describes the equilibrium and thus also the pressure of aluminum. For these conditions, the temperature for the onset of corundum condensation is 1758 K.

Chemical thermodynamics has been very successful in explaining the composition of meteorites in the Solar System. Many of the minerals in meteorites follow the condensation sequence of a slowly cooling Solar nebula (Grossman and Larimer, 1974). Likewise, the measured composition of presolar dust isolated from meteorites with an isotopic composition betraying an origin in the ejecta from late-type stars—so-called stardust—follows the thermodynamic condensation sequence (Lodders & Fegley, 1999). Observationally, infrared spectra of dust around such objects also reveal the relevance of the condensation sequence. This latter record is illustrated in **Figure 2**. Astronomical observations show that silicate and oxide dust form readily in the ejecta of oxygen-rich (elemental abundance of O > elemental abundance of C) Asymptotic Giant Branch stars for mass loss rates in the range $10^{-7} - 10^{-4}$ $M_\odot$/yr. The physical conditions around the sonic point—where the bulk of the dust must condense—in these winds can be estimated from mass and momentum considerations (**Figure 2**). At the lower mass loss rates (≃ $10^{-7}$ $M_\odot$/yr), the newly formed dust is observed to be dominated by oxides such as $Al_2O_3$ and MgFeO, but for mass loss rates exceeding some $10^{-6}$ $M_\odot$/yr amorphous silicates dominate the dust budget (Sloan and Price, 1998; Cami, 2001; Posch et al., 2003). The IR spectra of AGB stars with the highest mass loss rates (∼ $10^{-4}$ $M_\odot$/yr) show evidence for crystalline forsterite and enstatite grains. It seems thus that, as the mass loss rate increases, the calculated thermodynamic condensation sequence is 'followed' to lower temperatures, possibly reflecting the importance of freeze-out in the rapidly expanding AGB shells for lower mass loss rates (Tielens et al., 1997; Cami, 2001).

Thermodynamics should be used with care as chemistry is kinetics and thermodynamics, at best, only indicates the direction chemistry tends to flow. Often, reaction barriers are involved and then timescales can become very long compared to kinetic timescales of the system under consideration. Consider again the fiducial reaction (1). Based on quantum chemical calculations (Gobrecht et al., 2022), the chemical routes involved in the formation of the monomer and the dimer of corundum are shown in **Figure 4**. As this figure illustrates, AlO is the main aluminum-bearing species involved and extra oxygen that is required is provided by $H_2O$ and OH, where the latter is part of the (kinetic) $H-H_2-O-OH-H_2O$ system. In thermodynamics none of this matters as everything is regulated by (measured or calculated) Gibbs free energies. But in reality, many of the relevant reactions may have activation barriers and therefore be frozen out at the temperatures and pressures involved. Abundances of key species are then much lower than expected. By the same token, the solid surface may first have to be activated before reaction can occur and that too may be inhibited by an activation barrier. We will come back to this latter point when discussing carbon soot formation (**Section 3**).

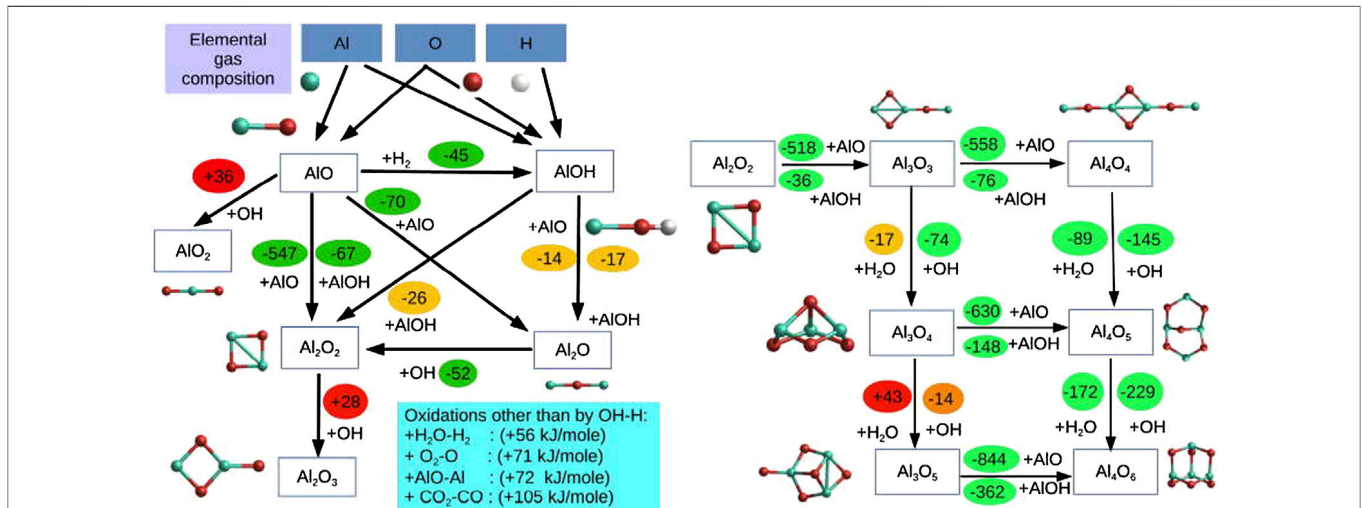

**FIGURE 4** | Reaction scheme for the formation of the alumina monomer (left) and dimer (right). Reaction enthalpies are given in kJ mol$^{-1}$. Very exothermic reactions are colour-coded in green, slightly exothermic reactions in orange, and endothermic reactions in red. The molecular structures of the aluminum-bearing species are indicated (blue is aluminum, red is oxygen). Figure taken from Gobrecht et al. (2022).

As **Figure 4** illustrates, attaining chemical thermodynamic equilibrium is inhibited by nucleation. The first steps in the condensation process involves nucleation of small clusters (monomers, dimers, … ) from simple gas phase species. The Gibbs free energies involved in cluster formation are in general different from those determined from atomization of solids. For clusters, these Gibbs free energies can be determined from quantum chemistry. Based on formalisms developed for the growth of raindrops in the atmosphere, this is often caught by adding a macrosocopic quantity—surface free energy—to the Gibbs free energy: $\Delta G(r) = \Delta \mu \, 4\pi r^3/3 + \gamma \, 4\pi r^2$ with $r$ the size of the cluster. As $\Delta \mu$ is negative, the volume term drives the growth while the positive surface term describes the free energy penalty of the surface. It is easy to see then that there is critical size, $d\Delta G/dr = 0$, which corresponds to a barrier to growth. Clusters smaller than this size will tend to dissolve while larger clusters will grow. In a slowly cooling gas, condensation will occur at a temperature well below the condensation temperature. In classical nucleation theory, a macroscopic approach is formulated to what is essentially a kinetic process. Really, the chemical routes involved in nucleation should be identified (see **Figure 4**) and these may, for example, contain activation barriers that are not caught by a macroscopic surface free energy term (e.g., in the formation of the radicals AlO and OH).

There is a second point hidden in these schematic thermodynamic schemes (**Figure 2**). Some of the reactions involve solid phase equilibria. For solid forsterite in contact with silica melt, reaction can proceed towards the left upon cooling.

Equivalently, upon heating, an enstatite crystal will segregate into a forsterite crystal plus silica melt. However, under typical conditions in space, melts are not relevant, and reaction would have to occur between forsterite and gaseous SiO (e.g., out of equilibrium) or a forsterite grain would have to interact with a silica grain and collisions between grains might be very rare.

The presence of $Mg_{0.1}Fe_{0.9}O$ grains in some AGB outflows (**Figure 2**) may be another manifestation of the importance of kinetics in astronomical dust condensation environments. While metallic iron is the thermodynamically favored compound under these conditions (**Figure 1**), because of nucleation issues, it is expected to form only under highly supersaturated conditions (Kozasa & Hasegawa, 1988). As metallic iron condensation is delayed, Fe might be incorporated into magnesium silicate or condense out as oxides. Observationally, circumstellar silicates are highly magnesium-rich (Fe/Mg < 0.01; Molster et al. (2002); Blommaert et al. (2014)) The 19 $\mu$m band observed in some AGB IR spectra is evidence for the presence of $Mg_{0.1}Fe_{0.9}O$, a solid solution of periclase and wuestite (Posch et al., 2003). Thermodynamically, wuestite is the favored condensate at high temperatures under limited (reducing) conditions. At low temperatures ( < 900 K), FeO is metastable as the conversion process to metallic iron is very slow (Liu et al., 2014). Smoke condensation experiments on Ca-Fe-Si-H$_2$-O$_2$ mixtures also reveal the fundamentally non-equilibrium condensation process with, among others, the formation of calciowuestite grains (Rietmeijer et al., 2009).

Finally, we have to consider that when kinetics is involved, time is of the essence. Consider the key reaction in astrophysics,

$$H_2 + H_2 \leftrightarrow H + H + H_2. \qquad (4)$$

The rate for this equation is $k(\,(4)) = 3.7 \times 10^{-10} \exp[-48350/T]$ cm$^3$ s$^{-1}$. **Figure 5** provides the timescale as a function of temperatures for this reaction at three different pressures. The two lowest pressures are particularly relevant for the inner solar nebula or for inner protoplanetary disks. The higher pressure is more relevant for exoplanet atmospheres. For stellar ejecta, with a dynamical timescale of ~ 10$^4$ yr, the atomic H fraction can be appreciable. The O/OH/$H_2$O fraction is set by the H/$H_2$ fraction and OH (and O) will burn carbonaceous solids to CO and $CO_2$. In the end, all of this will occur then on time scales set by the H/$H_2$ reaction.

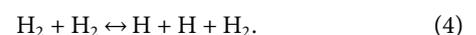

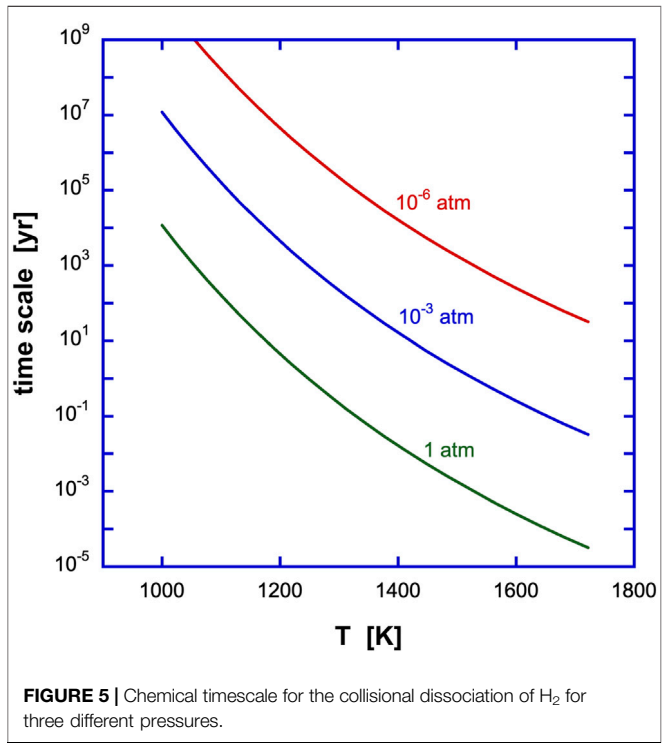

**FIGURE 5 |** Chemical timescale for the collisional dissociation of $H_2$ for three different pressures.

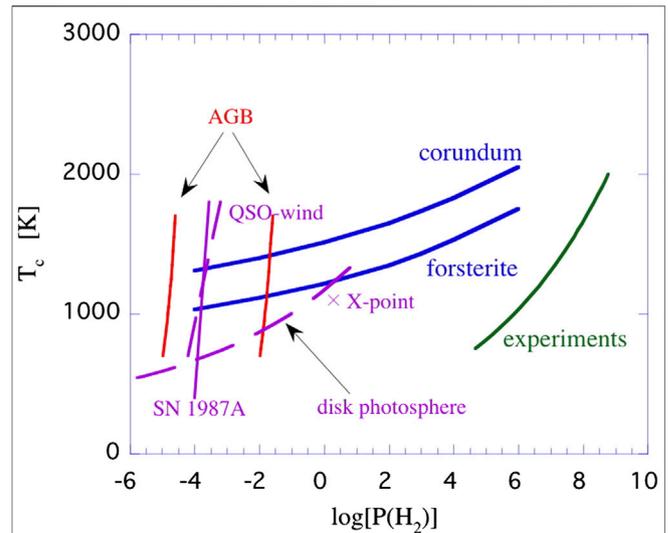

**FIGURE 6 |** Theoretical, experimental and astronomical studies on the condensation of silicates and oxides in a gas with Solar elemental composition. Total hydrogen pressures are in dyn cm$^{-2}$. The solid curves labelled corundum and forsterite represent the results of thermodynamic equilibrium calculations for the condensation temperature at a given pressure (Salpeter, 1977) (see also **Figure 1**). The results of laboratory studies on the critical pressure at which nucleation (of $Si_2O_3$) was detected in a SiO gas at a given temperature is shown as the green line labeled experiments (Nuth & Donn, 1982). These critical SiO pressures have been converted into $H_2$ pressures using the Si elemental abundance. The two almost vertical red lines labelled AGB show the relation between pressure and temperature in the circumstellar winds of Asymptotic Giant Branch stars around the sonic point for two different mass loss rates (10$^{-7}$–10$^{-4}$ $M_\odot$/yr). Because of freeze out, for these dust-driven winds, the bulk of the dust should condense close to the sonic point. The solid purple line indicates the estimated gas pressure and temperature at day 615 in SN 1987A when dust formation was clearly proceeding (Wooden et al., 1993). The top part refers to the gas temperature in the C-O zone cooled by gaseous CO (~1800 K), while the bottom part is appropriate for the observed dust temperature (400 K). The dashed purple line labeled QSO refer to conditions expected for the quasar wind in PG 212 + 059 where presumably newly condensed oxides have been detected in the mid-IR spectrum (Markwick-Kemper et al., 2007). Calculated temperatures and pressures in the photosphere of a passively heated disk at a distance of 0.1-1 AU around a T-Tauri star are shown as a dashed purple line (Chiang & Godreich, 1999). For such an externally irradiated disk, temperatures in the interior are much less. Actively accreting disks will have much higher temperatures in their mid-planes (as well as in their photospheres). The X indicates the estimated conditions at the X-point in magnetocentrifugally driven flows from T-Tauri stars.

A variety of laboratory studies on the condensation of dust have been performed, mainly focussing on the physical properties of the condensed grains (Nuth & Donn, 1982; Hallenbeck et al., 1998; Fabian et al., 2000; Demyk et al., 2001; Nuth et al., 2002; Rotundi et al., 2002). The results for the critical SiO pressure at which nucleation takes place in an SiO-$H_2$ gas (Nuth & Donn, 1982) are compared to the thermodynamic equilibrium calculations in **Figure 6**. These critical pressures are displayed as equivalent $H_2$ pressures adopting the Solar elemental Si abundance. This assumes implicitly that $H_2$ does not actively participate in (the rate-limiting step of) the nucleation process, which is borne out by experiments. These experimental pressures for the onset of nucleation exceed the thermodynamic equilibrium calculations by many orders of magnitude. It is clear that non-thermodynamic considerations must play a key role in these experiments. Likely, at the short timescale for nucleation in these experiments (~ 10 s), the kinetics associated with the chemical pathways towards the critical nucleation clusters dominates. As a corollary, the experiments may have only limited value in terms of characterizing condensation products.

Timescales seem to be much less of an issue for astronomical settings. Nevertheless, kinetic considerations are undoubtedly important. In recent years, a detailed study of condensation in O-rich environments has been started because of its interest in stardust studies, supernova dust condensates—particularly in the early Universe –, and dust formation in exoplanetary atmospheres (Cherchneff & Dwek, 2010). These studies bear out the importance of kinetic considerations. The role of kinetics in aluminum oxide cluster formation has been described above (c.f., **Figure 4**). Similar quantum chemical studies have been performed on silicon carbide, titanium oxide, and forsterite clusters and implemented in models for dust formation in stellar ejecta (Gobrecht et al., 2016, 2017; Boulangier et al., 2019).

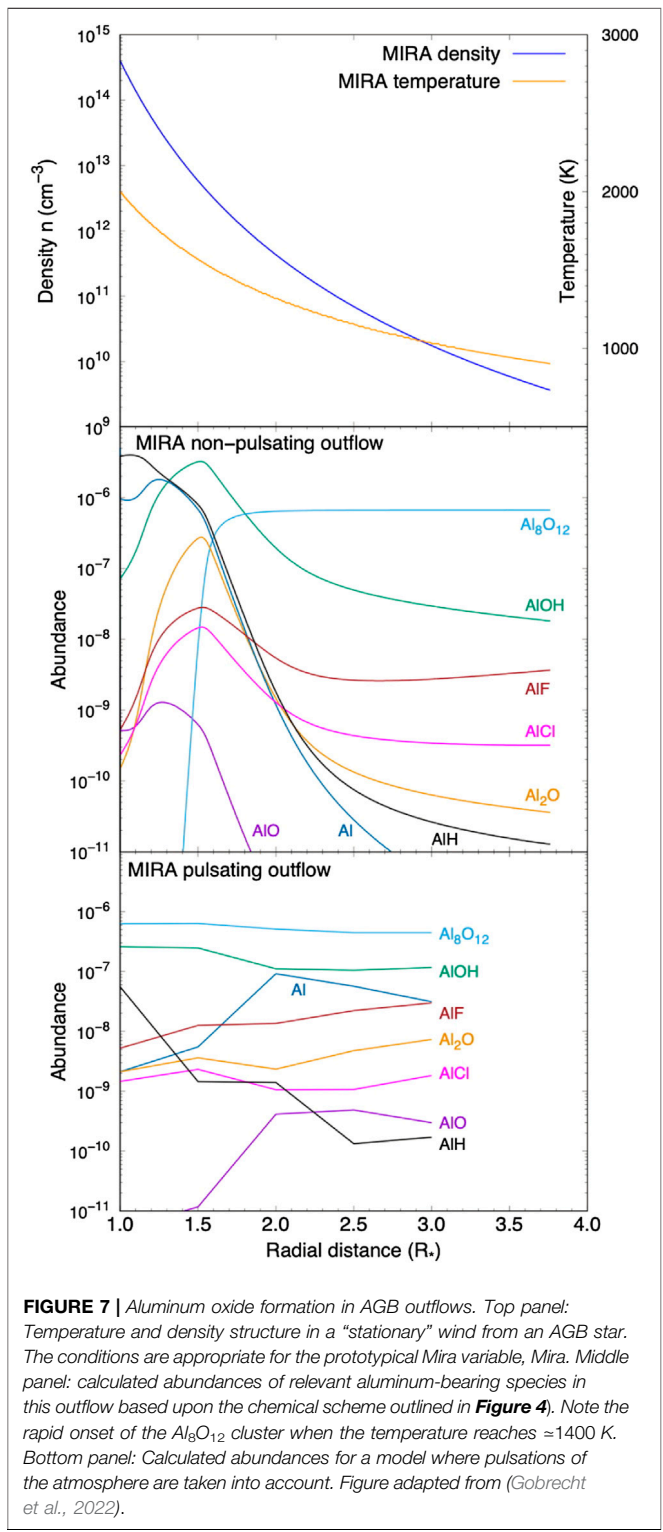

**FIGURE 7 |** *Aluminum oxide formation in AGB outflows. Top panel: Temperature and density structure in a "stationary" wind from an AGB star. The conditions are appropriate for the prototypical Mira variable, Mira. Middle panel: calculated abundances of relevant aluminum-bearing species in this outflow based upon the chemical scheme outlined in* **Figure 4***). Note the rapid onset of the $Al_8O_{12}$ cluster when the temperature reaches $\simeq$1400 K. Bottom panel: Calculated abundances for a model where pulsations of the atmosphere are taken into account. Figure adapted from (Gobrecht et al., 2022).*

When implementing these chemical routes, realistic models for astrophysical environments have to be considered. This is illustrated for aluminum cluster formation in **Figure 7**. The winds from AGB stars are accelerated by radiation pressure on dust. This is generally modeled as a radiation pressure force acting on a dusty fluid which is momentum-coupled to the gas (Tielens, 1983). This creates a time-independent (so-called stationary) structure where the velocity increases monotonically from the condensation layer through the sonic point to a typical value of 10–20 km/s. This acceleration plus geometric dilution leads then to a rapidly decreasing density structure. The temperature of the gas is set by a balance between dust-gas frictional heating and adiabatic expansion and molecular cooling, resulting in a temperature structure characterized by a modified adiabatic index (Tielens, 1983; Rodgers & Glassgold, 1991). The structure of such a model for the AGB star, Mira, are shown in **Figure 7**. Once the structure of the wind is determined, the results can be post-processed using a time-dependent chemical kinetic model that takes all relevant reactions into account. As **Figure 7** illustrates, with increasing distance from the photosphere, aluminum is converted into aluminium hydroxide, AlOH and then rapidly locked-up in the aluminum oxide tetramer. Indeed, the onset of aluminum oxide clusters is very rapid once the temperature reaches $\simeq$ 1400 K and the chemical network is driven to aluminum oxide clusters.

In reality, the structure of the Mira winds is more complex. The pulsations in these long-period variables drive strong shock waves into the photosphere. The shock compresses and heats the gas while it moves material outward but after the shock has passed, the material will fall back to almost its starting position while it cools down until the next shock passes by. In this way, material is slowly lifted away from the photosphere to a layer where dust condensation can take place. These recurring excursions in temperature and density can leave an imprint on the chemical composition of the gas. This can be modeled by following the Lagrangian trajectory of a gas parcel through its shock compression and heating and adiabatic expansion and cooling phases with a chemical kinetics network. This approach—pioneered by Cherchneff and coworkers (Cherchneff, 2006)—has now been applied to the formation of molecules and dust formation in oxygen-rich and carbon-rich AGB ejecta (Cherchneff, 2011, 2012; Decin et al., 2008; Gobrecht et al., 2022). **Figure 7** shows abundances calculated after one complete pulsation as a function of distance and efficient cluster formation in the wind occurs already at one stellar radius. As this study illustrates, the shock wave leaves the gas out of equilibrium and that greatly promotes dust nucleation. With ALMA and the VLT, observations of the composition of the dust condensation layers has come into reach and these can be compared to the results of such model studies. The first studies of this type are very encouraging (Decin et al., 2008, 2017, 2018; Kamiński et al., 2016, 2017; Takigawa et al., 2017).

Finally, we should mention the concept of heterogeneous nucleation. The first, high temperature dust compounds—such as $Al_2O_3$ in oxygen-rich AGB stars or titanium carbide in carbon-rich environments—will nucleate and then react on with gaseous species (**Figure 2**). These first grains may form the condensation nuclei for the main condensates—magnesium-rich silicates and carbon soot –; thereby circumventing the nucleation bottleneck. This may greatly facilitate the chemical growth of dust in stellar ejecta, but the chemistry involved in the first steps has not been studied in detail yet. Individual graphite stardust grains isolated from meteorites often show the presence of small TiC grains near their center, implying that this process is of importance in stellar ejecta (Bernatowicz et al., 1991, 2005; Vollmer et al., 2009; Singerling et al., 2021).

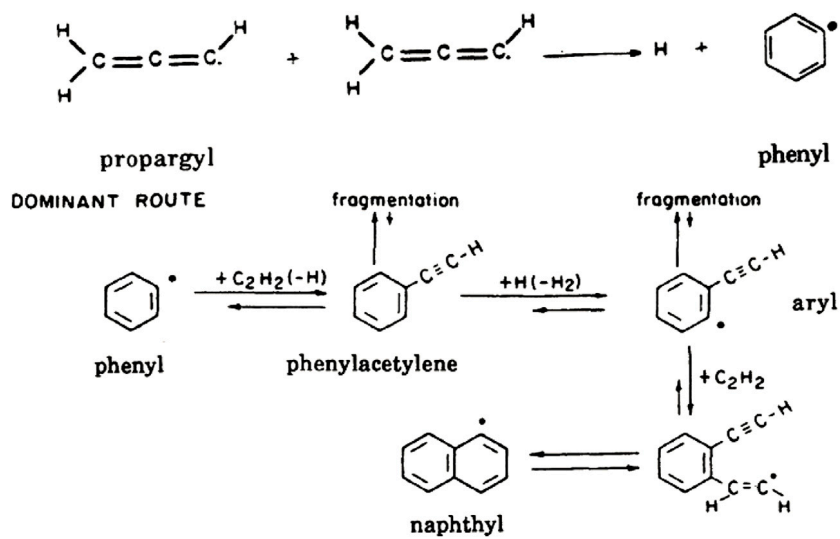

**FIGURE 8 |** *Top: One important route toward closure of the first ring. The reaction of two propargyl radicals forming the phenyl radical (C₆H₅). Bottom: Further growth consists of sequential steps of radical-site formation followed by acetylene addition, followed by ring closure. Figure adapted from Cherchneff et al. (1992).*

## 3 CARBON DUST

### 3.1 Chemistry of Sooting Flames

Based upon the extensive literature for soot chemistry in terrestrial environments—largely driven by internal combustion engine efficiency, polution, and health hazard concerns –, the chemical pathways towards carbon dust have been modeled in an astrophysical setting, using detailed reaction networks (Frenklach & Feigelson, 1989; Cherchneff et al., 1992). Here, we will briefly review the kinetic steps toward carbon soot. Several different chemical pathways towards soot have been proposed based upon neutral radicals, ions, polyacetylene chains or fullerenes and each of these may dominate in different environments as different factors drive the system to the thermodynamically most favored state. Their high stability, the high tendency of aromatic fuels to soot, and the structural similarity between the carbon backbone of PAHs and soot, have lead to the notion that PAHs are building blocks—the condensation nuclei—of soot particles. Indeed, C-sooting flames typically produce ∼ 5 % of the available C in small PAHs.

Acetylene is the feedstock for carbon dust formation and the rate-limiting step is the closure of the first ring (Wang & Frenklach, 1997). The resonantly stabilized radical, propargyl ($C_3H_3$), may play an important role in the formation of the first ring (**Figure 8**). Propargyl itself is formed through the reaction of small hydrocarbon radical, $CH_2$, with acetylene, where $CH_2$ itself is formed through attack of H atoms on acetylene. Other reactions, e.g., 1,3-butadienyl (CHCCHCH) radical reacting with acetylene, can also be important routes toward the phenyl radical. Once the first ring has formed—the nucleation step –, further growth occurs via the Hydrogen Abstraction Carbon Addition (HACA) mechanism. Here, an H atom is abstracted from an aromatic ring. The resulting aryl radical reacts with acetylene. This is followed by another step of H abstraction forming a radical site that reacts with acetylene. Ring closure leads then to a PAH with an additional aromatic ring. This is illustrated in **Figure 8** for the formation of naphthalene from benzene (Wang & Frenklach, 1997). Reactions of the aryl radical with other species (e.g., diacetylene) may also lead to ring growth, but these species are typically less important as feedstock because of abundance considerations. Growth is then favored to occur through the sequence of pyrene, coronene, … This results in a "window of opportunity" for rapid PAH growth, 900 < T < 1100 K. For temperatures above 1100 K, both the HA and the CA reactions are reversible and entropy does not favor growth. At 1100 K, the CA reaction becomes irreversible but the HA is still reversible, driving rapid PAH growth. Below 900 K, the HA reactions also freeze out and chemical growth stops (Frenklach & Feigelson, 1989; Cherchneff et al., 1992). At this point, the third stage of grain growth becomes dominant where PAHs agglomerate into larger structures due to weak van der Waals bonding. This is growth in size but not in mass of condensed material.

### 3.2 Models for Soot Formation in C-Rich AGB Stars

The first models for carbon soot formation in stellar ejecta faced difficulties in condensing a large fraction of the available $C_2H_2$, essentially, because the flow through the 900–1100 K condensation zone is very fast in stationary AGB wind models (Frenklach & Feigelson, 1989; Cherchneff et al., 1992). Dust condensation efficiencies in agreement with observations are obtained when the effects of recurrent shocks are included. These cycle a parcel of gas many times through the nucleation and HACA-growth windows as it slowly moves outwards (Cherchneff, 2012).

This discussion focused on the carbon soot growth process in sooting flames that also fits well with the conditions in AGB winds. Experiments have revealed that there is a second carbon soot formation window that opens up at higher temperatures, ≃ 3500 K (Jäger et al., 2009). The chemical route(s) involved are presently unclear but this carbon soot is highly disordered and has a fullerene-like structure, consisting of bent graphene layers.

Under the conditions of carbon-rich AGB winds, soot formation is driven by the presence of H-atoms that create the radical sites in the HACA mechanism. However, if there are other efficient mechanisms for radical site creation, then carbon soot growth can take place under other conditions. UV photons might provide one source of radical formation. The carbon skeleton of PAHs is very robust to carbon loss. UV fragmentation studies have shown that the H-loss channel opens up earlier than the $C_2H_2$ loss channel (Jochims et al., 1994). The H-coverage of a PAH is then set by a balance of unimolecular UV fragmentation and H-addition reactions (Andrews et al., 2016) and, in the interstellar medium, in regions where hydrogen is mainly molecular but there are still penetrating UV photons, reactions of the radicals with hydrocarbon molecules can lead to PAH growth even at low temperatures. Experiments have shown that naphthyl radicals react rapidly with acetylene and vinyl acetylene even at low temperatures, through a submerged barrier (Parker et al., 2012, 2015).

# 4 CONCLUDING REMARKS

The diversity of stardust formation sites, each characterized by different physical conditions, composition, and evolutionary timescales leads to a bewildering array of compounds injected into the interstellar medium. This diversity has been characterized through astronomical observations of dust signatures and through detailed laboratory studies of stardust grains. In a way, these environments provide a "natural" laboratory for the study of dust condensation.

Over the last 25 years, in depth studies of this chemical diversity have come into play through observations of the infrared signatures of dust in such environments. The James Webb Space Telescope will take these types of studies to new arenas as the composition of interstellar and circumstellar dust can be probed in, for example, low metallicity environments and supernovae. At the same time, the Very Large Telescope and the Atacama Large Millimeter Array can probe the chemical composition of the condensation zone in these stellar ejecta. The next generation of extremely large telescopes will be able to probe this in greater spectral and spatial detail.

Laboratory studies of individual stardust grains reclaimed from meteorites provide a different and unique view of the composition and formation history of dust under astrophysical conditions. Analytical techniques have been perfected to isolate such grains from meteorites and interplanetary dust particles and study them in detail—in composition, morphology and structure. In many ways, taking the ashes of stars apart one atom at a time can be considered a triumph of modern science and allows us to unravel the nucleosynthetic processes taking place in the fiery cauldrons of stellar cores. For this audience, though, the detailed view this provides of dust formed in well-defined stellar ejecta is a valuable guide to the processes in exoplanetary atmospheres.

Observations have revealed that dust particles form in the atmospheres of exoplanets and brown dwarfs and that this has profound influence on their thermal structure, chemical composition, and emerging spectra. It is not the purpose of this paper to review these aspects. Rather, the reader is referred to comprehensive reviews on the observational aspects as well as the physical and chemical processes that play a role in these environments (Helling, 2019, 2021; Gao et al., 2021). There are similarities and differences with the conditions for circumstellar dust formation. The elemental composition of hot Jupiters and brown dwarfs are very similar to those of O-rich AGB stars with highly reducing conditions and a prevalence of molecules, while supernova and nova ejecta may be relevant for exoplanet atmospheres where H is depleted. For the upper parts of the atmosphere, where dust formation will be relevant, temperatures are also very similar. However, the pressures in these atmospheres are typically higher than for circumstellar media. Comparison of $P - T$ profiles for exoplanetary atmospheres with thermodynamic condensation diagrams (c.f., **Figure 1**), would indicate that dust formation occurs in the range $T \sim 1000$–1500 K and $P \sim 10^{-5} - 10^{\circ}$ atm. Indeed, there are observational indications that dust formation may already set in at pressures of $10^{-4}$ atm in exoplanet atmospheres. Nevertheless, these pressures are many orders of magnitude higher than in circumstellar ejecta (**Figure 6**). Such high pressures will shorten chemical timescales by five orders of magnitude or more. However, in exoplanet atmospheres, dynamical timescales—atmospheric circulation and gravitational settling timescales—are measured in hours and are hence still very short compared to chemical timescales. Indeed, only for temperatures above 1500 K might the H/$H_2$ abundance be in equilibrium at 1 atm (**Figure 5**) and this ratio sets the ratio of many other species including OH/$H_2$O, which is key to oxidation. Similar to the shock wave treatment of AGB ejecta, proper linking of chemical kinetics to eddy circulation will then be important to follow the nucleation and chemical growth steps in exoplanet atmospheres.

One key lesson from circumstellar dust studies is that thermodynamics can provide insight into the chemical compounds formed in different environments. However, kinetics is of great importance as freeze out of reactions regulates the dust "condensation" window" and thereby the compounds formed. The oxide-silicate sequence in long-period-variables provides a telling example. Timescales for circumstellar dust formation range from tens to hundreds of days in supernova and novae explosions to years in AGB and red supergiant winds. Chemical nucleation is generally the rate limiting step. While classical nucleation theory is relevant for the formation of raindrops where molecules are bound by weak van der Waals forces, dust formation in astrophysical environments requires the identification of the chemical routes towards the formation of the first clusters and the reactions that allow their growth. These initial steps often involve radicals whose abundance is by necessity low and limits nucleation and growth. As emphasized above, while pressures and timescales are different, kinetic effects can be expected to be as relevant for dust formation in exoplanet atmospheres as for stellar ejecta. Hence, progress in this field will require the identification of relevant chemical routes toward dust nucleation and growth. In this sense, observations of circumstellar dust and laboratory analysis of stardust may be of great assistance for studies of dust in exoplanet and brown dwarf atmospheres.

Understanding these processes requires laboratory studies of relevant reaction rates and their branching ratios. Over the last decade, quantum chemistry studies have also matured enough to provide the necessary insight in these detailed chemical schemes. Laboratory dust condensation studies can provide insight in the structure of compounds formed. However, the difference in timescales between laboratory experiments (minutes) and astrophysical environments always has to be kept in mind in assessing their relevance.

## AUTHOR CONTRIBUTIONS

AT is solely responsible for the content of this article.

## FUNDING

Studies of the physics and chemistry of the interstellar medium at Leiden University are supported by a Spinoza award from the Dutch Science Foundation, NWO.

## ACKNOWLEDGMENTS

Dr. David Gobrechts kindly provided **Figure 7**.

**Conflict of Interest:** The author declares that the research was conducted in the absence of any commercial or financial relationships that could be construed as a potential conflict of interest.

**Publisher's Note:** All claims expressed in this article are solely those of the authors and do not necessarily represent those of their affiliated organizations, or those of the publisher, the editors and the reviewers. Any product that may be evaluated in this article, or claim that may be made by its manufacturer, is not guaranteed or endorsed by the publisher.